\documentclass[12pt]{article}
\usepackage{amssymb}
\usepackage{epsfig}
\newcommand{\be}{\begin{equation}}
\newcommand{\ee}{\end{equation}}
\newcommand{\bea}{\begin{eqnarray}}
\newcommand{\eea}{\end{eqnarray}}

\begin{document}

\begin{center}
{\bf$ (\beta \beta)_{0\,\nu}$-decay: a possible test of the
nuclear matrix element calculations}
\end{center}

\begin{center}
S. M. Bilenky\\
\vspace{0.3cm}
{\em IFAE, Facultat de Ciencies, Universitat Autonoma
de Barcelona, 08193, Bellaterra, Barcelona, Spain \\}

\vspace{0.3cm} {\em  Joint Institute    for Nuclear Research,
Dubna, R-141980, Russia\\}
\end{center}
\begin{center}
J. A. Grifols\\
\vspace{0.3cm}
{\em IFAE, Facultat de Ciencies, Universitat Autonoma
de Barcelona, 08193, Bellaterra, Barcelona, Spain \\}

\end{center}

\begin{abstract}
The existing calculations of the nuclear matrix elements of the
neutrinoless double $\beta$-decay differ by about a factor three.
This uncertainty prevents quantitative interpretation of the
results of experiments searching for this process. We suggest here
that the observation of the neutrinoless double $\beta$-decay of
{\em several} nuclei in future experiments could allow to test
different calculations of the nuclear matrix elements through the
direct comparison of them with the experimental data.
\end{abstract}

The compelling evidences in favor of neutrino oscillations, driven
by small neutrino masses and mixing, were obtained  recently in
the atmospheric \cite{AS-K,Soudan/MACRO} and in the solar
\cite{Cl,S-K,SNO,SNONC,SNOCC} neutrino
experiments. The small neutrino masses are generally considered as
a signature of a new scale in physics. There are no doubts that
the recent experiments are only the beginning of a long period of
investigations in this new field.

The investigation of neutrino oscillations, in which flavor lepton
numbers are not conserved, can not answer the question of the {\it
nature} of neutrinos with definite masses ({\it Dirac} or {\it
Majorana}?). The answer to this fundamental question can be
obtained by the study of processes in which the total lepton
number L is not conserved.

The most sensitive process to the violation of L and to the small Majorana neutrino masses is
the neutrinoless double $\beta$ -decay
($ (\beta \beta)_{0\,\nu}$-decay).

At present, data of the several experiments on the search for $
(\beta \beta)_{0\,\nu}$ - decay are available (see
\cite{Cremonesi}). In these experiments no indications on favor of
$ (\beta \beta)_{0\,\nu}$ - decay were obtained. \footnote{ The
recent claim \cite{Klap01} of some evidence of the $ (\beta
\beta)_{0\,\nu}$-decay, obtained from the reanalysis of the data
of the Heidelberg-Moscow experiment, was strongly criticized in
\cite{FSViss02/bb0nu02}.}

We will discuss $ (\beta \beta)_{0\,\nu}$-decay in the minimal scheme of  three-neutrino mixing \footnote{ We will not consider the data of the LSND experiment \cite{LSND}.
The LSND result will be checked
by the MiniBooNE
experiment \cite{MiniB}, started recently. }
\be
\nu_{lL} = \sum_{i=1}^{3} U_{li} \nu_{iL}\,.
\label{001}
\ee

Here $U$ is the unitary mixing matrix, and $\nu_{i}$ is the field
of the Majorana neutrino with mass $m_{i}$. The field
$\nu_{i}$ satisfies the Majorana condition
$$\nu_{i}=\nu^{c}_{i}= C\,\bar \nu^{T}_{i}\,,$$

where $C$ is the charge conjugation matrix. 

The matrix element of the $ (\beta \beta)_{0\,\nu}$ - decay
is proportional to the effective Majorana mass

\be
<m> = \sum_{i}^{3} U^{2}_{ei}\,m_{i}.
\label{002}
\ee

The elements $U_{ei}$ are complex. Let us write down \be U_{ei}=
|U_{ei}|\,~e^{i\,\alpha_{i}}\,. \label{005} \ee In the case of the
CP-invariance in the lepton sector, we have \cite{BNP}

\be U^{*}_{ei} =\eta^{*}_{i}\,U_{ei}\,, \label{006} \ee where
$\eta_{i}= i\,\rho_{i}$ is the CP-parity of the neutrino $\nu_{i}$
($\rho_{i}=\pm 1$). From (\ref{005}) and (\ref{006}) we find that
in the case of the CP-invariance

$$ 2 \alpha_{i} = \frac{\pi}{2}\,\rho_{i}$$

The strongest limits on
$|<m>|$ were obtained from the Heidelberg-Moscow  \cite{HM}
and IGEX \cite{IGEX} experiments.
In the Heidelberg-Moscow experiment  the following bound
for the lifetime of the  $ (\beta \beta)_{0\,\nu}$ - decay of
$^{76} \rm{Ge}$ was found

\be
T^{0\nu}_{1/2}(^{76} \rm{Ge})\geq 1.9 \cdot 10^{25}\, y\,~~ (90\% \,\rm{CL})\,.
\label{003}
\ee
Taking into account the results of the different calculations of the nuclear matrix elements,
from (\ref{003})
it was obtained
$$|<m>|\leq (0.35-1.24)\,~\rm{eV}\,.$$

Similar bounds were obtained in the IGEX experiment

\be
T^{0\nu}_{1/2}(^{76} \rm{Ge})\geq 1.57 \cdot 10^{25}\, y\,~~ (90\%\,\rm{CL})); \,
|<m>|\leq (0.33-1.35)\,~\rm{eV}\,.
\label{004}
\ee

Several new experiments on the search for $ (\beta \beta)_{0\,\nu}$ - decay
are in preparation and under development at present \cite{Cremonesi}.
In the experiments CUORE \cite{CUORE}, GENIUS \cite{GENIUS} , MAJORANA
\cite{MAJORANA},
EXO \cite{EXO} and MOON \cite{MOON} the sensitivities 

$|<m>|\simeq 2.7\cdot10^{-2}\,\rm{eV}$, $1.5\cdot10^{-2}\,\rm{eV}$,
$2.5\cdot10^{-2}\,\rm{eV}$, $5.2 \cdot10^{-2}\,\rm{eV}$ and
$3.6\cdot10^{-2}\,\rm{eV}$, respectively, are planned to be reached. \footnote{ In the calculation of these
sensitivities the nuclear matrix elements, given in \cite{Staudt}, were used.}

We will label neutrino masses in such a way that

$$ m_{1}< m_{2}< m_{3}$$
and denote neutrino mass squared differences as follows:
$$\Delta m^{2}_ {ik}=m^{2}_ {i}-m^{2}_{k}.$$
In the framework of the three-neutrino mixing all existing
solar and atmospheric neutrino data are described
in the leading approximation by two decoupled two-neutrino oscillations, which are characterized
by the parameters $\Delta m^{2}_{\rm{sol}}$, $\tan^{2}\theta_{\rm{sol}}$
and $\Delta m^{2}_{\rm{atm}}$, $\sin^{2}2\, \theta_{\rm{atm}}$
(see \cite{BGG}).

From the analysis of the Super-Kamiokande atmospheric neutrino
data it was found \cite{AS-K} 
\be 
1.6 \cdot 10^{-3} \leq \Delta
m^{2}_{\rm{atm}}\leq 3.9 \cdot 10^{-3} \rm{eV}^{2}\,~~
(90\%\rm{CL}) 
\label{010} 
\ee 
The best fit value of $\Delta m^{2}_{\rm{atm}}$ is equal to 
\be 
\Delta m^{2}_{\rm{atm}}= 2.5
\cdot 10^{-3} \rm{eV}^{2}\,~(\chi^{2}_{\rm{min}}=163/170\,\rm{dof})\,.
\label{011} 
\ee
 From the analysis of all solar neutrino data,
including the data of the Super-Kamiokande \cite{S-K} and
SNO \cite{SNO,SNONC,SNOCC} experiments, it was found that the most
plausible solution of the solar neutrino problem is the MSW LMA
solution. In \cite{SNONC}  the following ranges  were
found for the oscillation parameters:
 \be 
2.2 \cdot 10^{-5} \leq \Delta
m^{2}_{\rm{sol}}\leq 2.0 \cdot 10^{-4} \rm{eV}^{2}\,~ 0.22\leq
\tan^{2}\theta_{\rm{sol}}\leq 0.59\,~(99.7\%\rm{CL})\,.
\label{012} 
\ee

The best fit values of the parameters are equal to

\be \Delta m^{2}_{\rm{sol}}= 5.0 \cdot 10^{-5}
\rm{eV}^{2};\,~~\tan^{2}\theta_{\rm{sol}}= 0.34\,~
(\chi^{2}_{\rm{min}}=57/72 \,\rm{dof})\,. \label{013}
\ee

 The neutrino oscillation experiments do not allow to distinguish two
possibilities:

$$\Delta m^{2}_{21} \simeq \Delta m^{2}_{\rm{sol}};\,~
\Delta m^{2}_{32} \simeq \Delta m^{2}_{\rm{atm}}\,,$$

which corresponds to the normal hierarchy of neutrino mass-squared differences
($\Delta m^{2}_{21}\ll \Delta m^{2}_{32}$ ) or

$$\Delta m^{2}_{21} \simeq \Delta m^{2}_{\rm{atm}};\,~
\Delta m^{2}_{32} \simeq \Delta m^{2}_{\rm{sol}}\,,$$

which corresponds to the inverted hierarchy of neutrino mass-squared 
differences
($\Delta m^{2}_{21}\gg \Delta m^{2}_{32}$) .

The effective Majorana mass
$|<m>|$ depends on the type of the neutrino mass spectrum and
on the values of parameters
 $m_{1}$, $\Delta m^{2}_{\rm{atm}}$, $\Delta m^{2}_{\rm{sol}}$,
$\tan^{2}\theta_{\rm{sol}}$, $|U_{e3}|^{2}$ and on Majorana phase
differences. The element $|U_{e3}|^{2}$ is small as can be
inferred from the reactor long baseline experiments CHOOZ
\cite{CHOOZ2} and Palo Verde \cite{PaloV}. At the Super-Kamiokande
best fit value $\Delta m^{2}_{\rm{atm}}= 2.5\cdot
10^{-3}\rm{eV}^{2} $ we have from the CHOOZ exclusion curve \be
|U_{e3}|^{2}\leq 4\cdot 10^{-2}\,~~ (95 \%\rm{CL}) \label{014} \ee

The effective Majorana mass $|<m>|$ was calculated under different
assumptions about the neutrino mass spectrum in many papers (see
\cite{BPP} and references therein). Below we will discuss some
important implications of the measurement of the quantity $|<m>|$
for the neutrino mixing.

\begin{enumerate}

\item

The plausible
mechanism of the neutrino mass generation is the see-saw mechanism \cite{see-saw}.
 If neutrino masses are of the see-saw origin, they satisfy
 the hierarchy 

\be 
m_{1}\ll m_{2}\ll m_{3}\,. 
\label{016} \ee
Hence for the effective  Majorana neutrino mass we have in the
see-saw case
$$
|<m>|\leq \sin^{2}\,\theta_{\rm{sol}}\,\sqrt{\Delta m^{2}_{\rm{sol}}}+ |U_{e3}|^{2}\,\sqrt{\Delta m^{2}_{\rm{atm}}}\,,$$
Using the best-fit values of the solar neutrino oscillation parameters and
the CHOOZ bound
(\ref{014}) for the effective Majorana mass
 we obtain the upper bound

$$
|<m>| \leq 3.8 \cdot 10^{-3}\,\rm{eV}\,,$$

which is significantly smaller than the expected sensitivity
of the future experiments.
Thus,
if  $ (\beta \beta)_{0\,\nu}$ - decay
is observed in the experiments of the next generation 
it could be an argument against the neutrino mass hierarchy (\ref{016})
and 
the standard see- saw mechanism.

Let us notice that in the case of the inverted mass hierarchy
$$m_{1}\ll m_{2}<m_{3}$$

 we have the range
\be \sqrt{\Delta m^{2}_{\rm{atm}}} \,~
|\cos2\,\theta_{sol}|\leq|<m>| \leq \sqrt{\Delta
m^{2}_{\rm{atm}}}\, \label{017} \ee for the effective Majorana
mass.

 Using the best fit values of the oscillation parameters,
from (\ref{017}) we obtain

$$
2.5 \cdot 10^{-2}\,\rm{eV} \leq |<m>|\leq 5 \cdot
 10^{-2}\,\rm{eV}\,.$$

Thus, in the case of the inverted mass hierarchy the predicted value
of $|<m>|$ is in the range of the sensitivity of the future
neutrinoless double $\beta$ -decay experiments.
This is connected with the fact that
the size of $|<m>|$ is determined in this case by the
``large'' $\sqrt{\Delta m^{2}_{\rm{atm}}}$.
In the case of the normal mass hierarchy the contribution of $\sqrt{\Delta m^{2}_{\rm{atm}}}$
to the effective Majorana mass is suppressed by the smallness
of $|U_{e3}|^{2}$.

\item
The best laboratory upper bound \footnote{At this point we should
remark that existing large scale structure surveys combined with
CMB data give \cite {Elgaroy} : $\sum m_{i} \le
(1.8-2)\,~\rm{eV}\,~~~( 95\%\,~\rm { CL})\,$. } on the absolute
value of the neutrino mass was obtained from the tritium
$\beta$-decay experiments Mainz \cite{Mainz} and Troitsk
\cite{Troitsk}:

$$m_{\beta}< 2.2\,~\rm{eV}\,~~~( 95\%\,~\rm { CL})\,.$$

A new tritium experiment on the measurement of the neutrino mass
KATRIN \cite{Katrin} is currently in preparation. In this experiment
one expects to reach the sensitivity

$$m_{\beta}< 0.35\,~\rm{eV}\,.$$

If the neutrino mass $m_{\beta}$ is measured in the KATRIN experiment,
it would mean that neutrino masses are practically degenerate and
due to the unitarity of the mixing matrix
$$m_{\beta}\simeq m_{1} .$$

For the effective Majorana mass, independently of the type
of the mass spectrum,
we will have in this case
$$ |<m>| \simeq m_{\beta}\,~|\sum_{i=1}^{3}U_{ei}^{2}|.$$

Neglecting small contribution of
$|U_{e3}|^{2}$ ($|U_{e1}|^{2}$ in the case of the inverted hierarchy),
we can connect the parameter $\sin^{2}2\,\alpha $
($\alpha =\alpha_{2}-\alpha_{1}$ or $\alpha =\alpha_{3}-\alpha_{2}$)
with  measurable quantities. We have \cite{BGP}

\be
\sin^{2}\alpha \simeq \left( 1 -\frac{ |<m>|^{2}}{m_{\beta}^{2}}
\right)\,~\frac{1}{\sin^{2}2\,~\theta_{\rm{sol}}}\,.
\label{018}
\ee

If CP is conserved
$\sin^{2}\alpha =0$ ( $\sin^{2}\alpha =1$)
for the case of the same (opposite) CP parities of
$\nu_{1}$ and  $ \nu_{2}$ (or $\nu_{2}$ and  $ \nu_{3} $).
Thus, the measurement of the effective Majorana mass
in the future $ (\beta \beta)_{0\,\nu}$ experiments and the neutrino mass
$m_{\beta}$ in the tritium KATRIN experiment
could  allow to obtain an information on the CP violation
in the lepton sector.

\item

We have discussed the cases of small and large $m_{1}$.
Now we will consider the case of the intermediate $m_{1}$:
$$m_{1}\sim\sqrt{\Delta m^{2}_{\rm{atm}}}.$$

The investigation of the $ (\beta \beta)_{0\,\nu}$ - decay could
be an important source of information about the minimal mass
$m_{1}$ in this case. In fact, in the case of the normal hierarchy we
have

\be |<m>|\simeq m_{1} \,\left( 1 -
\sin^{2}2\,\theta_{\rm{sol}}\,\sin^{2}\,\alpha\right)^{1/2}\,,
\label{019} 
\ee

From this expression for $m_{1}$ we obtain the range

\be
|<m>|\leq m_{1}\leq \frac{|<m>|}{|\cos2\,\theta_{\rm{sol}}|}\,,
\label{020}
\ee

In the inverted hierarchy case we have

\be
|<m>|\simeq \sqrt{m_{1}^{2}+\Delta m^{2}_{\rm{atm}}}
\,\left( 1 - \sin^{2}2\,\theta_{\rm{sol}}\,\sin^{2}\,\alpha\right)^{1/2}\,,
\label{021}
\ee

For the minimal mass we obtain from (\ref{021})

\be
\sqrt{||<m>|^{2}-\Delta m^{2}_{\rm{atm}}|}\leq m_{1}
\leq\sqrt{\frac{|<m>|^{2}}{\cos^{2}2\,\theta_{\rm{sol}}}
 -\Delta m^{2}_{\rm{atm}}}
\label{022}
\ee

\end{enumerate}

Thus, the measurement of the effective Majorana mass
$|<m>|$ could reveal those properties of the neutrino masses and mixing that
can not be investigated in the neutrino oscillation experiments or
in the tritium experiments.

There is, however,
a problem with the determination of $|<m>|$  from the experimental data.
The total probability of the
 $ (\beta \beta)_{0\,\nu}$ - decay has the following general form

\be
\Gamma^{0\,\nu}(A,Z) = |<m>|^{2}\,|M(A,Z)|^{2}\,G^{0\,\nu}(E_{0},Z)\,,
\label{023}
\ee
where $ M(A,Z)$ is the nuclear matrix element and
$G^{0\,\nu}(E_{0},Z)$
is a known phase space factor ($E_{0}$ is the energy release).
In order to determine  $|<m>|$ from the experimental data we need
to know the nuclear matrix element. This last quantity must be calculated.

There exist at present two basic approaches to the
calculation of the $ (\beta \beta)_{0\,\nu}$ nuclear matrix elements: the
quasiparticle random phase approximation and the nuclear shell model (see
\cite{Elliott} and references therein).
The results of different calculations of the
lifetime of the $ (\beta \beta)_{0\,\nu}$-decay differ
by about one order of magnitude.
For example, if we assume that $|<m>|= 5\cdot 10^{-2}\,\rm{eV}$,
from the results of different calculations of
the nuclear matrix element for the lifetime of the $(\beta \beta)_{0\,\nu}$ - decay of $^{76}\rm{Ge}$
we will obtain the values in the interval \cite{Elliott}:

$$6.8\cdot 10^{26}\leq T^{0\,\nu}_{1/2}(^{76}\rm{Ge})\leq 70.8\cdot 10^{26}\,\rm{years}$$

The problem of the calculation of the nuclear matrix elements
of the neutrinoless double $\beta$-decay is a real theoretical challenge.
Without its solving the effective Majorana neutrino mass can not
be determined with reliable accuracy
from the results of $ (\beta \beta)_{0\,\nu}$ experiments.

Uncertainties of the nuclear matrix elements of
$ (\beta \beta)_{0\,\nu}$-decay were discussed recently by the authors
of ref. \cite{Barger} in their discussion of the
possibility
to study CP violation in $ (\beta \beta)_{0\,\nu}$-decay.

We would like to propose here a possible method which allows to test
the results of the calculation of the nuclear matrix elements
of the neutrinoless double $\beta$-decay. We will take into account
the following

\begin{enumerate}

\item A few eV sensitivity
is planned to be reached in
experiments searching for neutrinoless double $\beta$ -decay
of {\em different} nuclei:
$^{76}\rm{Ge}$, $^{136}\rm{Xe}$, $^{130}\rm{Te}$, $^{100}\rm{Mo}$
and others;

\item The effective Majorana mass $ |<m>|$
is determined only by the Majorana neutrino masses and elements
of the neutrino mixing matrix. For small neutrino masses
($m_{i}\lesssim 10\,\rm{MeV}$) the nuclear matrix elements do not
depend on neutrino masses \cite{Doi}.

\end{enumerate}

Thus, the ratio of the lifetimes of the $(\beta\,\beta)_{0\,\nu}$-decay of
different nuclei
depend only on nuclear matrix elements and known phase space factors.
If  the neutrinoless double $\beta$ -decay of different nuclei
is observed, the calculated ratios of the corresponding
lifetimes can be be confronted with the experimental values.
For illustration we present in the Table I
the ratios of lifetimes of the $(\beta\beta)_{0\,\nu}$-decay
of different nuclei, calculated in six
models.
For the lifetimes we used the  values given in \cite{Elliott}. As it is seen from Table I, the calculated ratios
vary
by about one order of magnitude.

\begin{center}
 Table I
\end{center}
\begin{center}
The results of the calculation of the ratios of the lifetime of
$(\beta\beta)_{0\,\nu}$-decay for different nucleus. The references
to the corresponding papers are given.
\end{center}

\begin{center}
\begin{tabular}{|ccccccc|}
\hline
Lifetime ratios
&
\cite{HS84}
&
\cite{Caurier99}
&
\cite{EVZ88}
&
\cite{Staudt}
&
\cite{TS95}
&
\cite{Pantis96}
\\
\hline
$T^{0\,\nu}_{1/2}(^{76}\rm{Ge})$/$T^{0\,\nu}_{1/2}(^{130}\rm{Te})$
&
11.3
&
3
&
20
&
4.6
&
3.6
&
4.2
\\
$T^{0\,\nu}_{1/2}(^{76}\rm{Ge})$/$T^{0\,\nu}_{1/2}(^{136}\rm{Xe})$
&

&
1.5
&
4.2
&
1.1
&
0.6
&
2
\\
$T^{0\,\nu}_{1/2}(^{76}\rm{Ge})$/$T^{0\,\nu}_{1/2}(^{100}\rm{Mo})$
&

&

&
14
&
1.8
&
10.7
&
0.9
\\
\hline
\end{tabular}
\end{center}
 In conclusion, we have shown here that the observation of the neutrinoless double $\beta$ - decay of $^{76}\rm{Ge}$, $^{136}\rm{Xe}$, $^{130}\rm{Te}$ and other nuclei in experiments of the next generation
will allow to confront the results of the different calculations of nuclear matrix elements of the $(\beta\,\beta)_{0\,\nu}$-decay directly with
experimental data.

S.M.B. acknowledges the support of the ``Programa de Profesores Visitantes de IBERDROLA de Ciencia y Tecnologia''.


\begin{thebibliography}{99}
\bibitem{AS-K} Super-Kamiokande Collaboration, S.~Fukuda {\it et al.,}
Phys. Rev. Lett. {\bf 81}, 1562 (1998);\,~
 S.~Fukuda {\it et al.,} Phys. Rev. Lett. {\bf 82}, 2644 (1999);\,~
S.~Fukuda {\it et al.,} Phys. Rev. Lett. {\bf 85}, 3999-4003 (2000).

\bibitem{Soudan/MACRO} Soudan 2 Collaboration, W.W.M.Allison \textit{et al.},
Physics Letters {\bf B 449} (1999) 137;  MACRO Collaboration,
M.Ambrosio et al. hep-ex/0106049; Phys. Lett. B517 (2001) 59 \,~M.
Ambrosio et al. NATO Advanced Research Workshop on Cosmic
Radiations, Oujda (Morocco), 21-23 March, 2001.

\bibitem{Cl} Homestake Collaboration B. T. Cleveland {\it et al.}, Astrophys. J. {\bf
496}, 505 (1998);
Kamiokande Collaboration,
Y. Fukuda {\it et al.}, Phys. Rev. Lett. {\bf 77}, 1683 (1996);
GALLEX Collaboration, W. Hampel
{\it et al.}, Phys. Lett. {\bf B 447}, 127 (1999);
GNO Collaboration,
M. Altmann {\it et al.}, Phys. Lett. {\bf B 490}, 16 (2000);
SAGE Collaboration,
J. N. Abdurashitov {\it et al.}, Phys. Rev. {\bf C 60}, 055801 (1999).

\bibitem{S-K} Super-Kamiokande Collaboration, S.~Fukuda {\it et al.}, Phys. Rev. Lett. {\bf 86}, 5651 (2001).

\bibitem{SNO} SNO collaboration, Q.R. Ahmad {\it et al.}, Phys. Rev. Lett.
 {\bf 87}, 071301 (2001). 
\bibitem{SNONC}
SNO collaboration Q.R. Ahmad {\it et al.}, Phys.Rev.Lett.
{\bf 89}, 011301 (2002); nucl-ex/0204008. 
\bibitem{SNOCC}
SNO collaboration, Q.R.
Ahmad {\it et al.,} Phys.Rev.Lett {\bf 89}, 011302 (2002);
nucl-ex/0204009.


\bibitem{Cremonesi} O.\,Cremonesi, Proceedings of the
20th International Conference on Neutrino
                Physics and Astrophysics,
{\em Neutrino~2002\/}\,(Munich, Germany) May 25-30, 2002.



\bibitem{Klap01} H.V. Klapdor-Kleingrothaus \textit{et al.}, Mod. Phys. Lett.
{\bf 16} (2001) 2409.

\bibitem{FSViss02/bb0nu02} F. Feruglio, A. Strumia and F. Vissani, hep-ph/0201291. C.E. Aalseth \textit{et al.}, hep-ex/0202018.

\bibitem{LSND} LSND Collaboration, G. Mills, Proceedings of the
19th International Conference on Neutrino
                Physics and Astrophysics,
{\em Neutrino~2000\/}\,(Sudbury, Canada) June 16-21, 2000.

\bibitem{MiniB} MiniBooNE Collaboration, R. Tayloe,
Proceedings of the 20th International Conference on Neutrino
Physics and Astrophysics, {\em Neutrino~2002\/}\,(Munich, Germany)
May 25-30, 2002.
\bibitem{BNP} L. Wolfenstein,  Phys. Lett. \textbf{ B107} (1981) 77;\,~
 S.\ M. \ Bilenky, N.\ P.\ Nedelcheva and
S.\ T. \ Petcov, Nucl. Phys.  \textbf{ B247} (1984) 589;\,~
 B. Kayser, Phys. Rev. {\bf D30} (1984) 1023.
\bibitem{HM} HEIDELBERG-MOSCOW collaboration, H. V. Klapdor-Kleingrothaus
\textit{et al.}, Eur. Phys. J. \textbf{ A 12}, 147 (2001).

\bibitem{IGEX} IGEX Collaboration, C. E. Aalseth \textit{et al.},
Physics of Atomic Nuclei {\bf 63} (2000) 1225, hep-ex/0202026.

\bibitem{CUORE}  E. Fiorini, Phys. Rep. {\bf 307} (1998) 309.

\bibitem{GENIUS} H.\ V.\ Klapdor-Kleingrothaus \textit{et al.},
          J. Phys. \textbf{ G 24} (1998) 483.

\bibitem{MAJORANA} L. De Braeckeleer \textit{et al.}, Proceedings of
the Carolina Conference on Neutrino Physics,
           (Columbia (SC), USA) March 2000; C.E. Aalseth \textit{et al.},
hep-ex/0201021.




\bibitem{EXO}  M.\ Danilov \textit{et al.},
Phys.\ Lett. {\bf B480} (2000) 12.

\bibitem{MOON}
H. Ejiri {\em et al.}, Phys. Rev. Lett.
                         {\bf 85} (2000) 2917.

\bibitem{Staudt} A.\ Staudt, K.\ Muto, H.\, V.\, Klapdor-Kleingrothaus,
Europhys. Lett. {\bf 13} (1990) 31.


\bibitem{BGG} S. M. Bilenky, C. Giunti and
W. Grimus,  Prog. Part. Nucl. Phys. {\bf 43} (1999) 1;\,~ hep-ph/9812360.


\bibitem{CHOOZ2} CHOOZ Collaboration,  M.\, Apollonio \textit{et al.}, Phys.
Lett. {\bf B 466} 415 (1999).


\bibitem{PaloV} F.\, Boehm, J.\, Busenitz et al., Phys.\ Rev.\ Lett.  {\bf 84}, 3764 (2000);\,~
Phys. Rev. {\bf D 62} (2000) 072002.

\bibitem{BPP} S.M. Bilenky, S. Pascoli and S.T. Petcov, Phys. Rev.
{\bf D 64} (2001) 113003. S. Pascoli and S.T. Petcov,
hep-ph/0205022.





\bibitem{see-saw}
M.~Gell-Mann, P.~Ramond and R.~Slansky,
in \textit{Supergravity}, p.~315, edited by F. van Nieuwenhuizen and D.
  Freedman, North Holland, Amsterdam, 1979\,;
T.~Yanagida,
Proc. of the \textit{Workshop on Unified Theory and the Baryon Number of the
  Universe}, KEK, Japan, 1979\,;
R.N. Mohapatra and G.~Senjanovi{\'c}, Phys. Rev. Lett. \textbf{44}, 912 (1980).

\bibitem{Mainz} Ch. Weinheimer, Proceedings of
the 20th International Conference on Neutrino
Physics and Astrophysics, {\em Neutrino~2002\/}\,(Munich, Germany)
May 25-30, 2002.


\bibitem{Troitsk} V.\ Lobashev \textit{et al.}, Nucl. Phys. Proc. Suppl.
{\bf 91 }(2001) 280.

\bibitem{Elgaroy} O. Elgaroy \textit{et al.,} astro-ph/ 0204152.


\bibitem{Katrin} KATRIN collaboration, A.Osipowicz \textit{et al.},
hep-ex/0109033.

\bibitem{BGP} S. M. Bilenky, C. Giunti and  S. T. Petcov, Phys. Rev.
\textbf{D54} (1996) 4432.




\bibitem{Elliott} S. R. Elliott and P. Vogel, Annu. Rev. Nucl. Part.
 Sci. \textbf{52} (2002), hep-ph/0202264.

\bibitem{Barger} V. Barger, S. L. Glashow, P. Langacker and D. Marfatia,
hep-ph/0205290.

\bibitem{Doi} M. Doi, T. Kotani and E. Takasugi, Progr. Theor. Phys. Suppl.
 \textbf{53} (1985) 1.








\bibitem{HS84}W. C. Haxton  and  G. J. Jr.Stephenson, Progr. Part. Nucl.
Phys.\textbf{12} (1984) 409.

\bibitem{Caurier99} E. Caurier, F. Nowacki, A. Poves, J. Retamosa, Phys. Rev.  Lett. \textbf{77}(1996) 1954 ;
 E. Caurier, F. Nowacki, A. Poves, J. Retamosa, Nucl. Phys. A \textbf{A 654}
(1999) 973.

\bibitem{EVZ88} J. Engel, P. Vogel, M. Zirnbauer, Phys. Rev. C \textbf{37}
(1988) 731.

\bibitem{TS95}J. Toivanen and J. Suhonen, Phys. Rev.  Lett. \textbf{75}
(1995) 410.

\bibitem{Pantis96}G. Pantis, F. \v{S}imkovic, J. D. Vergados, A. Faessler,
 Phys. Rev. \textbf{C 53} (1996) 695.




\end{thebibliography}
\end{document}